\documentstyle[twocolumn,aps,epsfig,floats]{revtex}
\draft 

\begin{document} 
\twocolumn[\hsize\textwidth\columnwidth\hsize\csname @twocolumnfalse\endcsname
%==============================================================================
\title{
Formation of hyperfine fields in alloys
}
\author{
A. K. Arzhnikov and L. V. Dobysheva}

\address{
Physical-Technical Institute, Ural Branch of Russian Academy of Sciences, \\
Izhevsk, Russia
}
\maketitle

\begin{abstract}
This work deals with the analysis of experimental data on the
average magnetization of $Fe_{1-x}Me_x$ (Me=Sn,Si) disordered alloys, the
average and local hyperfine fields (HFF) at the Fe nuclei.  The effect of
the metalloid concentration on the HFF is studied with the help of the
results of first- principles calculations of ordered alloys. The disorder is
taken into account by means of model systems. The dependences obtained
correspond to those experimentally observed. Experimental data on the ratio
of the average HFF at Fe nuclei to the average magnetisation in alloys with
sp-elements show that the ratio decreases proportionally with the metalloid
concentration. This change in the ratio is bound up with three factors.
First, the contribution of the valence electron polarization by the
neighboring atoms, that is positive (unlike the polarization by the own
magnetic moment), increases with the change of the disorder degree (increase
of concentration). Second, the appearance of the impurities, i.e. metalloid
atoms, in the nearest environment of Fe leads to the orbital moment
increase. And, finally, the change of the disorder degree, as in the first
case, results in an increase in the orbital magnetic moment and its positive
contribution to the HFF. The value and the degree of the influence of these
contributions to the HFF is discussed.
\end{abstract}

]
\narrowtext  

In the experimental physics of solids there are few methods for measuring
local characteristics. Prominent among them are nuclear techniques (e.g.
Mossbauer spectroscopy) determining the hyperfine magnetic fields (HFF) at
nuclei. It is widely believed that these techniques can reflect some
chemical and topological features of the atomic surroundings of the excited
nuclei.  Indeed, in numerous studies these spectra are interpreted in terms
of phenomenological models considering only the nearest atomic environment
(see e.g. [1-3]). There is no doubt that such a description is often useful
and efficient. We think, however, that there are some cases when such a
simplified approach is not justified. Moreover, for the spectra of
transition metals and alloys on their basis, the successful description
within the framework of such models is rather an exception than a rule, and
every time additional arguments are needed to justify these limitations. For
example, such an approach makes some sense in the case of disordered systems
due to strong localization of the d- electrons responsible for the itinerant
magnetism of transition metals [4]. This is one of the reasons why sometimes
the spectra of disordered alloys are successfully described by the
Jaccarino-Walker- type models where the local magnetic moments (LMM) and
HFFs are assumed to be proportional to the number of metalloid atoms in the
nearest environment.

In our opinion, even in this case the experimental spectra may provide
more reliable and complete information without the use of
phenomenological models and restriction to the nearest environment. One
should then analyze the HFF peculiarities from the "first-principles"
calculations.  This work is devoted to such an analysis of the hyperfine
magnetic fields, magnetization and local magnetic moments and their
interrelation in the disordered $Fe_{1-x}Sn_x$, $Fe_{1-x}Si_x$ alloys.
Here the magnetic characteristics of clusters with a given impurity
configuration in disordered alloys are taken from the first-principles
calculations of the translationally invariant systems, supposing the
interaction between the clusters to be not significant because of small
free length of electron in these alloys. Nowadays, there is no
possibility to conduct these calculations for the disordered systems,
moreover, it should be noted that even calculations of the ordered
alloys often do not give the required agreement with experiment and
reveal only the main features of the HFF and LMM behavior.

The calculations were performed by the full-potential linearized augmented
plane wave method (FLAPW) using the WIEN-97 program package [5]. The results
are presented in Table 1.

The systems were simulated on a BCC lattice which in the disordered alloys
under consideration is retained within a wide concentration range [1]. The
lattice parameters were chosen in accordance with the experimental values
for x=3.125 at.\% and x=6.25 at.\%. It should be mentioned that even at small
concentrations the BCC lattice is somewhat distorted due to the
repulsion/attraction by the Sn/Si atom of the surrounding Fe atoms. As shown
in our paper [6], the changes in magnetic characteristics because of this
relaxation are insignificant. Though the results in Table 1 were obtained
with allowance for this relaxation, we do not discuss it here.

1. AVERAGE AND LOCAL MAGNETIC MOMENTS.

Fig.1 presents the experimental data and the calculated averages of the
magnetic moment per one Fe atom. 
%==============================================================================
\begin{figure}[bt]  
\epsfig{file=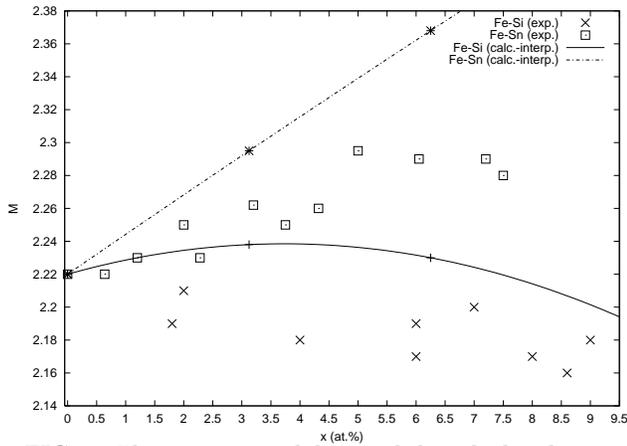,width=8.5cm}
  \caption{The experimental data and the calculated averages of the magnetic
moment per one Fe atom}
 \label{magn_mom}
\end{figure}
%==============================================================================
The average magnetic moment  $M=\sum_i
Md^i +M^{int}$, where $Md^i$ is the spin magnetic moment over the
muffin-tin (MT) sphere of the i-th Fe atom (hereinafter we will refer to
this value as local magnetic moment, LMM), $M^{int}$ is the spin
magnetic moment over the unit cell without the MT- spheres. The main
contribution to the LMM comes from the d-electrons that are almost
entirely inside the MT-sphere, whereas $M_{int}$ is formed by the s- and
p- electrons and has a small negative value as compared to the LMM. We
do not take into account the contribution of the orbital magnetic moment
that is about 0.045 $\mu_B$ in our calculations. In experiments this
value is 1.5?2 times larger and comprises ~0.08 $\mu_B$ [7]. It is
believed that the inclusion of the orbital polarization in the
exchange-correlation potential [8] improves the agreement between the
calculations and the experimental values, but here we didn't use such a
potential supposing that the calculated value can be multiplied by the
factor close to two. The main reason for the neglect of the orbital
contribution lies in the fact that its variations with concentration and
configuration of metalloid atoms are small as compared to M even with
allowance of the actual value and amounts less than 1 \% (see the last
but one column in Table 1), whereas the Mdi variations range up to 15
\%. Fig.1 displays a rather good agreement of the magnetic moment with
the experimental data, though the theoretical values of the magnetic
moment are somewhat higher than the experimental ones. We believe that
the disorder, which is not taken into account here, reduces the
magnetization by 2-3 \% [9]. Let us mention some peculiarities of the
magnetic moments formation.

A. The magnetic moment does not depend on the metalloid type and, as shown
in Ref.10, is governed by the lattice parameter, a. At equal concentrations,
the lattice parameter of the Sn alloy is greater than that of the Si alloy,
so the magnetic moment in the first case is greater.

B. The LMM of the Fe atom closest to the impurity is of the smallest among
others value (Table 1). As noted in Ref.10, this difference is defined by
the competition between two mechanisms: the LMM reduction due to flattening
of the d-band because of the s-d hybridization that is the strongest near
the impurity, and the LMM increase due to narrowing of the d-band because of
a decrease of the wave function overlap. There also exists third mechanism
of the LMM reduction due to the difference in the impurity-potential
screening by d-electrons (the difference in pushing out the impurity levels
by the bands with spin up and down), which was revealed in this work by a
comparison of the number of d-electrons within the MT-sphere for different
Fe-atom positions (there are more d-electrons near the impurity, Table 1).

C. The LMMs are concentration dependent. So the LMM of the Fe atom
closest to the Si atom is 2.181 $\mu_B$ at x=3.125 at.\% ($Fe_{31}Si$), and
2.262  $\mu_B$ at x=6.25 at.\% ($Fe_{15}Si$). This LMM increase with
concentration holds in general for all the non-equivalent positions
(Table 1). However, in spite of the LMM increase, the average magnetic
moment in the $Fe_{1-x}Si_x$ system somewhat decreases 2.238 $\mu_B$ ->
2.230 $\mu_B$ (Fig.1). This is the result of a higher probability of
finding a Fe atom with a neighboring impurity, which increases the
number of Fe atoms having lower LMM values.

2. HYPERFINE MAGNETIC FIELDS AT NUCLEI.

The program package WIEN-97 allows one to calculate the interaction
between the nucleus magnetic moment and the spin and orbital magnetic
moments of the electron subsystem. The spin dipole contributions are
small (~2 $\div$ 3 kGs), they are suppressed due to the symmetry
relations, and the main contribution comes from the spin polarization at
the nucleus (Fermi-contact interaction) and the orbital magnetic moment.
The Fermi-contact interaction may be divided into $H^{core}_i$ (the
core-electron polarization) and $H^{val}_i$ (the valence-electron
polarization), and hence, the resulted field at the i-th site contains
three terms: $H^i=H^{core}_i+H^{val}_i+H^{orb}_i$. For the core-electron
polarization the simple relation $H^{core}_i =\gamma^s Md_i$ is
satisfied, where $\gamma^s$ does not depend to a high accuracy on
neither the metalloid type nor its  concentration, and is determined
only by the approximation for the exchange-correlation potential.  Here
we use the GGA approximation [11] which gives $\gamma^s\approx 123 
kGs/\mu_B$. In our opinion just this simple  expression between the LMM
and Hcore makes it possible to use phenomenological models  neglecting
the effect of the atoms in the second, third, etc. coordination spheres.
Really, from Table 1  one can see that the main distinctions  in LMM at
a certain concentration are connected with the  presence or absence of
the metalloid atom in nearest environment. So, this also determines the 
variations in $H^{core}_i$. Though the experiments showed that the
proportionality between the magnetic  moment and HFF  is not as good as
we found, and the proportionality coefficient essentially  decreases
with concentration. As we shall see later this is connected with the
other two contributions  to the HFF.

The proportionality $H^{orb}_i =\gamma^{orb} M^{orb}_i$ is fulfilled in
a somewhat worse way, but still rather satisfactorily. Note, however,
that $\gamma^{orb}$ is positive and about five times larger in magnitude
than  $\gamma^s$.  Hence, if the changes of $M^{orb}_i$ affect the
magnetic moment only slightly, the changes in Horbi may amount to 20 kGs
even at low concentrations (see Table 1). The nature of these changes in
disordered alloys is discussed in more detail in Ref.12. Here we mention
only the main features. The $H^{orb}_i$ increases along with the
$M^{orb}_i$ increase with concentration. The $H^{orb}_i$ takes the
largest value at the atom closest to the metalloid atom. The increase of
$H^{orb}_i$ with allowance for the actual values of $M^{orb}_i$, that
are twice as large as the calculated one, comprises 15$\div$20 kGs even
at low concentrations as compared to that in pure Fe. On the strength of
the qualitative character of the last statement, we can say about the
tendency of the variations. Finally, as shown in [10], the orbital
contribution increases also with disorder (that is, with concentration).

The $H^{val}_i$ behaves in a more complicated way. This is primarily
associated with strong delocalization of the s- and p-like electrons
that interfere at sites with different magnetic and charge properties,
and therefore the $H^{val}_i$ behavior cannot be in fact quantitatively
predicted. However, we succeeded in revealing some qualitative
regularities supported by experimental evidence. First of all, we
analyzed the valence contribution using the simple functional dependence
of the magnetic moment screening in the RKKY (Ruderman - Kittel - Kasuya
- Yosida) theory, as it was done in Ref.9:  $H^{val}_i= A+B \sum_j M^d_j
cos(2k_fr/T+\phi)/r^3$. Such a simplified treatment of $H^{val}_i$ is
hardly justified in our case, but we hope to have determined the main
qualitative dependences. Solving the inverse task for the ordered
cluster of size 200 a.u. we receive the most probable values of A, B, T
and $\phi$ in the alloys $Fe_{15}Sn$ and $Fe_{31}Sn$. The corresponding
functions of $B cos(2k_fr/T+\phi)/r^3$ are shown in Fig.2. 
%==============================================================================
\begin{figure}[bt]  
\epsfig{file=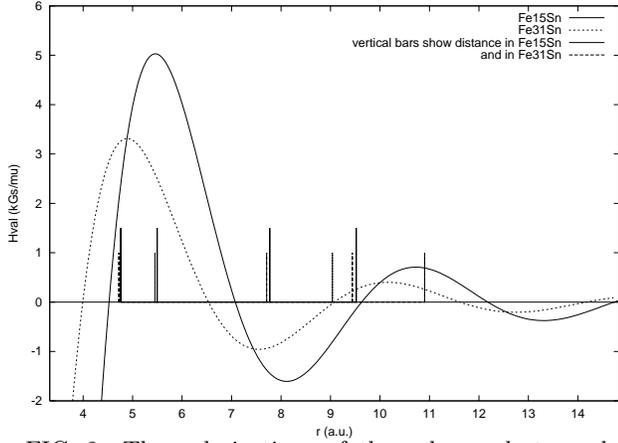,width=8.5cm}
  \caption{The polarizations of the valence electrons by a Fe-atom magnetic
moment as a function of distance}
 \label{RKKY}
\end{figure}
%==============================================================================
Of special
interest is the fact that for both concentrations the spin polarization
of electrons is positive for the I and II coordination spheres, which
entirely contradict the results of a similar processing of the
experimental data. This is due to the fact that during processing of the
experimental data the difference between H0 (HFF at the nucleus of the
Fe atom without the metalloid atoms in its nearest environment) and H1
(with one impurity atom in the nearest environment) was attributed to
the changes in  $H^{val}$, whereas in an alloy there are differences in
the local magnetic moment that primarily affect $H^{core}$ and 
$H^{orb}$. Fig.3 presents the experimental $H_0$ and $H_1$ as a function
of concentration.  
%==============================================================================
\begin{figure}[bt]  
\epsfig{file=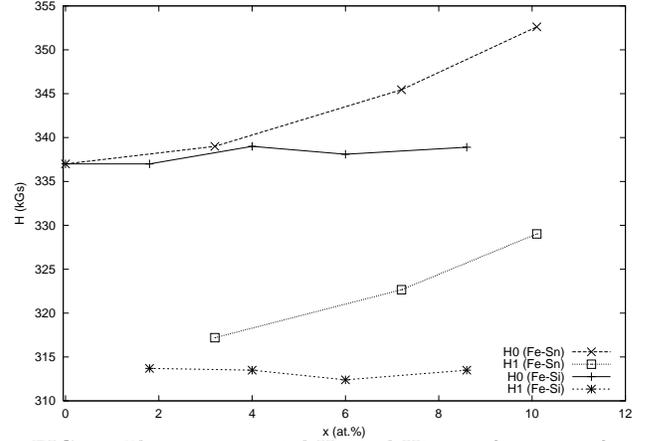,width=8.5cm}
  \caption{The experimental H0  and H1  as a function of concentration}
 \label{h0h1}
\end{figure}
%==============================================================================
Table 1 shows that the difference between these
quantities can be successfully explained by the LMM magnitude and hence
by the core-electron polarization. However, if everything were
determined only by  $H^{core}_i$ we should expect an increase in $H_0$
and $H_1$ with concentration in accordance with the LMM increase, which
is not the case, as for the $Fe_{1-x}Sn_x$ alloy the magnetic moment
increases much more quickly than H0 and H1, and there is no increase at
all for Si. In reality the expected increase is compensated by the
decrease in magnitude of the configurationally averaged negative
contribution $H^{val}$. Fig.4 gives the averaged values of $H^{val}$ of a
disordered cluster of size 200 a.u. with a certain number of impurities
in the first coordination sphere. 
%==============================================================================
\begin{figure}[bt]  
\epsfig{file=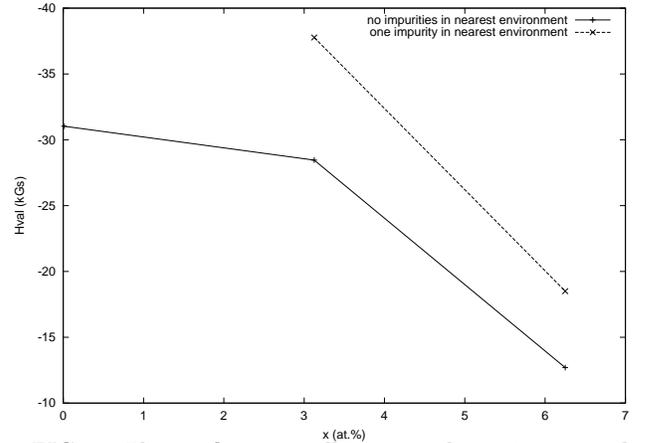,width=8.5cm}
  \caption{The configurationally averaged negative contribution Hval}
 \label{hval}
\end{figure}
%==============================================================================
The averaging was performed in assumption that the Fe atoms are
distributed randomly and polarize the conductivity electrons at distance
r according to the model function for concentrations 6.25 and 3.125 \%
(see Fig.2). The decrease of the averaged Hval with concentration is due
to the positive values of the RKKY polarization at the distance of the
first and second coordination spheres (see Fig.2). Two main features in
the behavior of the Hval averaged over configurations can be noticed.
First, the magnitude of $H_{val}$ at the Fe atom without impurities in
nearest environment is less by 5-10 kGs than $H_{val}$ at the Fe atom
with one impurity in the environment, and second, the magnitude of the
averaged  $H_{val}$ decreases with concentration for both configurations
of the environment.

Thus, the use of the "first-principles" calculations makes it possible
to explain the main peculiarities of the HFF behavior in the
low-concentration disordered alloys $Fe_{1-x}Me_x$. The difference
between $H_0$ and $H_1$  $H_0 - H_1\approx -20$ kGs consists of
$H^{val}_0 - H^{val}_1\approx 7\div 10$ kGs, $H^{core}_0 -
H^{core}_1\approx -15 \div - 25$ kGs and  $H^{orb}_0 - H^{orb}_1\approx 
-5 \div -10$ kGs. The decrease of the proportionality coefficient
between the HFF and the magnetic moment with concentration results from
the decrease of magnitude of $H_{val}$ and increase of magnitude of 
$H_{orb}$ that are opposite to the $H_{core}$.

We believe that the general relations obtained here will be useful in
processing the experimental data on the HFFs in disordered alloys of
transition metals and nonmagnetic impurities.

The present work was supported by Russian Foundation for Fundamental
Researches, Grant No 00-02-17355.

\newpage

\widetext
\begin{table}
\caption{The results of the calculations: Configuration of impurities in
the Fe-atom environment, [nm...] denotes the number of metalloid atoms
in the first (n), second (m) etc. spheres. Number of such Fe atoms in
the unit cell, $N_{Fe}$. Number of d-electrons in the MT sphere, $N^d$.
Magnetic moment, $M^d$, in the MT sphere.  Contribution of the
core-electrons polarization to the HFF, $H^{core}$. Contribution of the
valence-electrons polarization to the HFF, $H^{val}$. Orbital magnetic
moment, $M^{orb}$. Orbital contribution to the HFF, $H^{orb}$.}
\begin{tabular}{cccccccc}
   &$[nm...]$, $N_{Fe}$ & $N^d$ & $M^d$, $\mu_B$ & $H^{core}$,kGs 
& $H^{val}$,kGs & $M^{orb}$, $\mu_B$ & $H^{orb}$,kGs \\
\tableline
$Fe_{31}Sn$, & [1000], 8 & 6.006 & 2.246 & -277.50 & -37.59 & .0489 & 27.94\\
a=21.804a.u. & [0003], 8 & 5.951 & 2.507 & -310.35 & -30.94 & .0483 & 26.41\\
             & [0100], 6 & 5.964 & 2.432 & -300.17 & -34.84 & .0482 & 26.62\\
             & [0020], 6 & 5.963 & 2.425 & -299.79 & -22.39 & .0479 & 26.28\\
             & [0000], 2 & 5.987 & 2.391 & -295.13 & -34.34 & .0441 & 23.66\\
             & [0000], 1 & 5.959 & 2.423 & -299.68 & -24.17 & .0470 & 25.15\\
\tableline
$Fe_{15}Sn$, & [1003], 8 & 6.114 & 2.442 & -301.01 & -20.25 & .0548 & 29.77\\
2a=21.985a.u.& [0200], 3 & 6.090 & 2.530 & -312.64 & -29.97 & .0535 & 29.08\\
             & [0040], 3 & 6.084 & 2.542 & -314.55 &  -2.46 & .0550 & 29.38\\
             & [0000], 1 & 6.142 & 2.390 & -294.33 & -29.97 & .0450 & 23.32\\
\tableline
$Fe_{31}Si$, & [1000], 8 & 5.999 & 2.181 & -269.44 & -38.68 & .0497 & 28.27\\
a=21.604 a.u.& [0003], 8 & 5.967 & 2.413 & -297.89 & -35.15 & .0471 & 25.55\\
             & [0100], 6 & 5.977 & 2.343 & -288.51 & -31.31 & .0446 & 24.46\\
             & [0020], 6 & 5.965 & 2.385 & -294.06 & -16.92 & .0468 & 25.32\\
             & [0000], 2 & 5.965 & 2.489 & -308.13 & -19.72 & .0485 & 26.44\\
             & [0000], 1 & 5.969 & 2.389 & -294.81 & -13.56 & .0451 & 24.28\\
\tableline
$Fe_{15}Si$, & [1003], 8 & 6.130 & 2.262 & -277.20 & -29.11 & .0594 & 32.38\\
2a=21.585a.u.& [0200], 3 & 6.115 & 2.353 & -289.52 & -21.64 & .0456 & 24.44\\
             & [0040], 3 & 6.094 & 2.432 & -298.87 &   5.49 & .0509 & 27.31\\
             & [0000], 1 & 6.094 & 2.536 & -313.82 & -13.89 & .0496 & 26.83\\
\end{tabular}
\end{table}

\end{document}